\documentclass[a4paper,11pt]{article}

\usepackage{authblk}
\usepackage{amsmath}
\usepackage{amssymb}
\usepackage{url}
\usepackage{hyperref}
\usepackage{color,graphicx}
\usepackage{subcaption}
\usepackage{algorithmic}
\usepackage[ruled,lined,algo2e]{algorithm2e}

\usepackage{multirow}
\usepackage{mathtools}
\usepackage{lineno}
\usepackage{stmaryrd}

\usepackage{xcolor}

\newcommand\hl[1]{%
  \bgroup
  \hskip0pt\color{red!80!black}%
  #1%
  \egroup
}

\usepackage{adjustbox}

\newcommand{\claimqed}{\hspace*{\fill} $\triangle$  \ifmmode \else
    \par\addvspace\topsep\fi}

\newcommand\blfootnote[1]{%
  \begingroup
  \renewcommand\thefootnote{}\footnote{#1}%
  \addtocounter{footnote}{-1}%
  \endgroup
}

\begin{document}


\title{Spectrum graph coloring to improve Wi-Fi channel assignment in a real-world scenario via edge contraction}

\author[1]{David Orden}

\author[2]{Ivan Marsa-Maestre}

\author[2]{Jose Manuel Gimenez-Guzman}

\author[2]{Enrique de la Hoz}

\author[1]{Ana \'Alvarez-Su\'arez}

\affil[1]{Departamento de F\'{\i}sica y Matem\'aticas, Universidad de Alcal\'a, Spain.
\texttt{david.orden@uah.es}, \texttt{ana.alvarez@uah.es}}

\affil[2]{Departamento de Autom\'atica, Universidad de Alcal\'a, Spain.
\texttt{ivan.marsa@uah.es}, \texttt{josem.gimenez@uah.es}, \texttt{enrique.delahoz@uah.es}}

\date{}
\maketitle


\begin{abstract}
The present work deals with the problem of efficiently assigning Wi-Fi channels in a real-world scenario, the Polytechnic School of the University of Alcal\'a. We first use proximity graphs to model the whole problem as an instance of spectrum graph coloring, we further obtain a simplified model using edge contraction, and we finally use simulated annealing to look for a coloring which optimizes the network throughput. As the main result, we show that the solutions we obtain outperform the \emph{de facto} standard for Wi-Fi channel assignment, both in terms of network throughput and of computation time.
\end{abstract}


\blfootnote{\begin{minipage}[l]{0.2\textwidth} \includegraphics[trim=10cm 6cm 10cm 5cm,clip,scale=0.15]{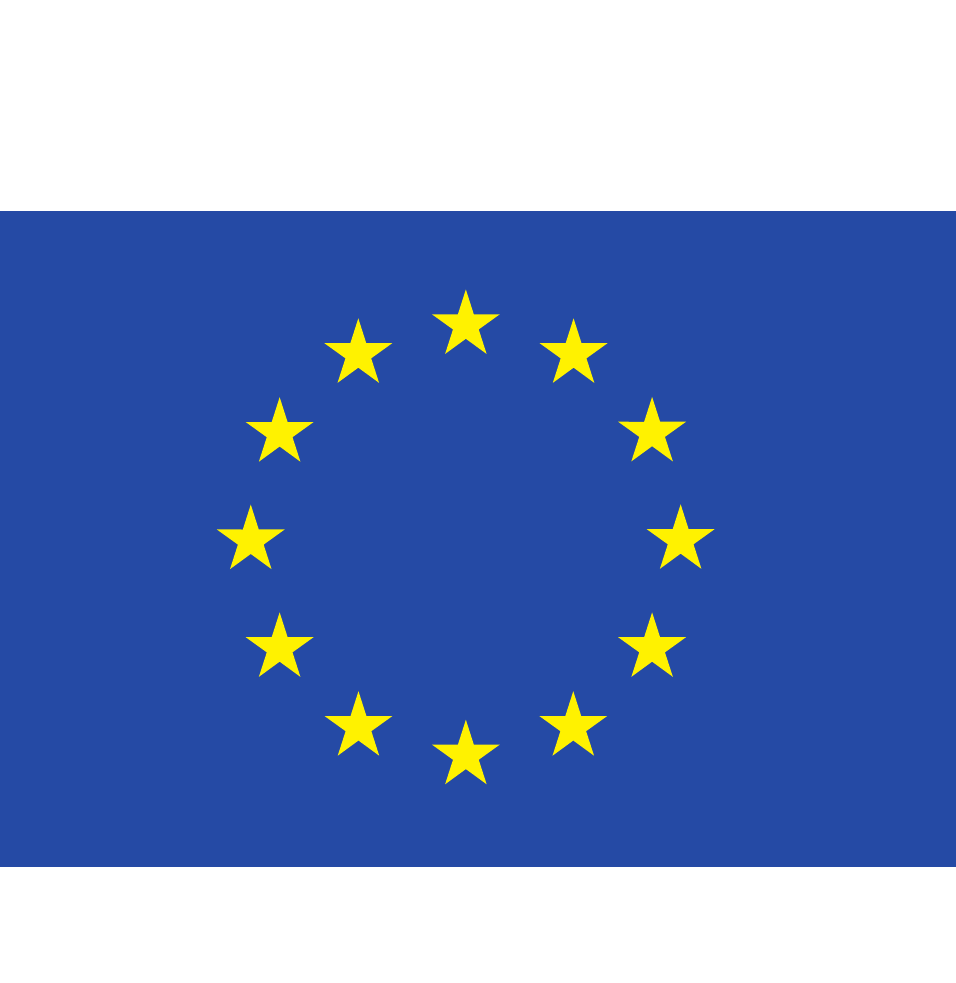} \end{minipage}  \hspace{-0.7cm}
\begin{minipage}[l][1cm]{0.8\textwidth}
 	  This work has received funding from the European Union's Horizon 2020 research and innovation programme under the Marie Sk\l{}odowska-Curie grant agreement No 734922.
 	\end{minipage}\vspace{-0.4cm}}



\section{Introduction}\label{sec:Introduction}

As wireless communications have become more and more widespread, frequency assignment problems (FAPs) have arisen. Although different models can be considered depending on the particular application, such as telephone and satellite networks, in general a frequency assignment problem can be described as
a set of connections to whom must be assigned frequency bands or channels, i.e, intervals of bandwidth from a given set, so that transmission is possible, and has some optimization criteria that take into account interference between channels~\cite{Aardal}. Interference occurs when two frequencies are close in the electromagnetic spectrum, or are harmonics, while also being in relatively close proximity, in such a way that one signal disturbs the other.

\begin{figure}[htb]

\includegraphics[width=\columnwidth]{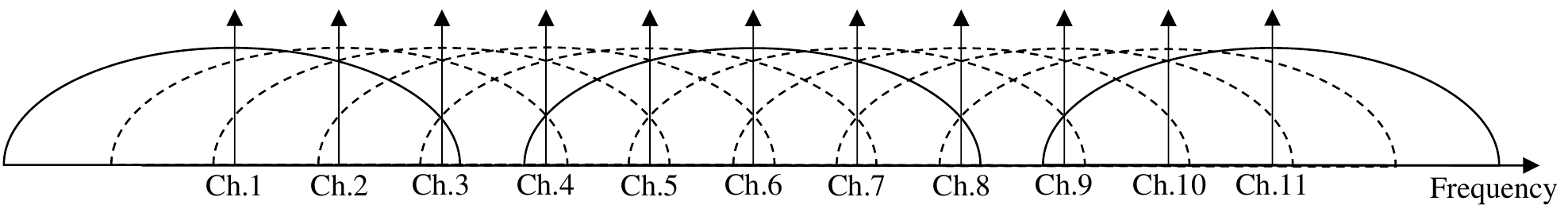}

\caption{Graphical representation of the Wi-Fi overlapping channels.}
\label{fig:wifichannels}
\end{figure}

A prominent case is the IEEE 802.11 (Wi-Fi) technology, which has become ubiquitous in our daily life, because of both its low cost and its use of unlicensed frequency bands. Most commonly, Wi-Fi operates in the 2.4 GHz wireless frequency band, which is divided into 11 frequency channels~\cite{Ng2012} which partially overlap, see Figure \ref{fig:wifichannels}. Each access point, e.g., a Wi-Fi router, is assigned to use one of these channels to operate.

Unfortunately, the aforementioned advantages of Wi-Fi are also a frequent cause of severe performance drawbacks.
The increasingly high number of Wi-Fi devices in use, added to the existence of other technologies and devices operating in the same band, translates into a great impact in performance.
It has been proved~\cite{Bazzi11} that the existing interference relates directly to the perceived throughput. Therefore, an accurate choice of the frequency channel used by each access point (AP) is essential to reduce interference and increase the wireless devices (WD) utility.

Different approaches have been considered for this problem of Wi-Fi channel assignment, which is becoming more and more relevant in the engineering community. We refer the reader to the comprehensive surveys~\cite{Aardal, Chieochan}. In this last survey~\cite{Chieochan} we can find a classification and the most prominent works regarding channel assignment in Wi-Fi networks. These techniques can be classified into \textit{centrally managed} and \textit{uncoordinated}. Moreover, as also pointed out in~\cite{Chieochan}, in this first category works follow two different perspectives depending whether they consider channel assignment together with AP placement or not. This distinction is key, as both problems are highly coupled and raise clearly different issues. In this work, we focus on assigning channels without considering AP placement, i.e., we consider that APs and their position are already in place. However, the interested reader can find in~\cite[Chapter 11]{koster2009graphs} some prominent advances in the area of joint solving AP placement and channel assignment.

Regarding the techniques that are usually used for solving the channel assignment problem, most of the proposals rely on heuristics, due to the complexity of the problem. However, optimization techniques have also been used~\cite{Seyedebrahimi2016}.

From the mathematical community, frequency assignment has traditionally been considered as one of the most prominent applications of graph coloring, with the well known vertex coloring problem modeling a scenario in which graph nodes are to be assigned a frequency and graph edges denote pairs of nodes in the range of interference, so that they must not be assigned the same frequency. Although this model, in which colors correspond to frequencies and hetero-chromatic edges guarantee the assignment of different frequencies to potentially interfering nodes, has been widely used, Tragos et al.~\cite{tragos2013spectrum} have established the necessity of a more accurate model to fit cases where interferences arise also between channels close enough in the spectrum. Of course, this implies a significant increase in complexity~\cite{Bar-Yehuda} and complicates the efficient computation of solutions. The book~\cite{koster2009graphs} is a comprehensive study of the use of graphs and algorithms in communication networks.

In this work, we present a model which finds an equilibrium between being complex enough to reflect all types of interferences, and being simple enough to allow an efficient computation of solutions. Our proposal stems from Spectrum Graph Coloring, a generalization of the classical vertex coloring problem which we have recently introduced~\cite{spectrumcoloring}. In this problem, we are given an abstract undirected graph $G$ and a spectrum of colors $S=\{c_1,\ldots,c_{s}\}$ endowed with a matrix~$W$
storing a non-negative distance $W_{ij}=W(c_i,c_j)$ between each pair of colors, so that these distances correspond to the interferences between Wi-Fi channels and, thus, a coloring $c$ of the graph induces at each vertex~$v$ an interference
\[I_v(G,W,c)=\sum_{u\in N(v)}W(c(u),c(v)).\]

\begin{figure}[!htb]
\begin{center}
\includegraphics[width=\textwidth]{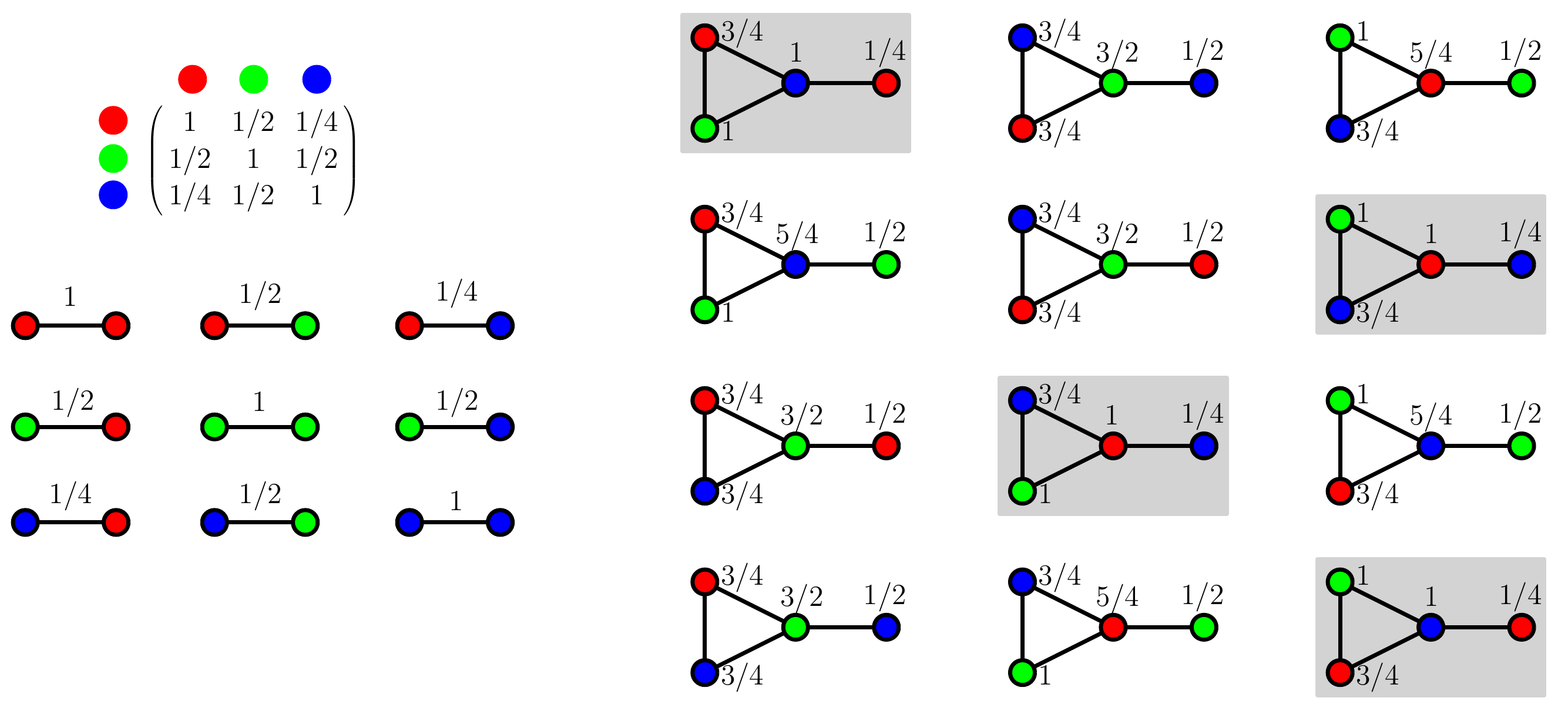}
\end{center}
\caption{Left: Matrix $W$ and the interferences it induces on the different possible colorings of an edge. Right: All proper 3-colorings of the paw graph and the interferences they induce at every vertex. Highlighted colorings achieve interference~$\leq 1$ at every vertex.} \label{fig:TotalInterferenceColoring}
\end{figure}

See an example in Figure~\ref{fig:TotalInterferenceColoring}, which shows a matrix~$W$ and all the proper 3-colorings of the paw graph together with the interferences they induce. Note that highlighted colorings achieve interference at most~$1$ at every vertex.

Hence, spectrum graph coloring adds to the extensive list of graph coloring problems. Many of them, including the $\lambda$-coloring~\cite{Bodlaender} and the \emph{distance coloring}~\cite{Sharp}, fit in the framework of $L(p_1,\ldots, p_k)$-labellings~\cite{Griggs}, where two vertices
separated by a distance $i$ must be labelled with colors whose mutual distance is at least~$p_i$. Some coloring problems have also arisen  in direct connection with the frequency assignment problem, such as \emph{$T$-colorings}~\cite{roberts-Tcolorings-91} and the \emph{$S$-packing coloring}~\cite{gastineau-Scoloring-15}. In the former, given a set $T$ of non-negative integers that represent disallowed separations between channels, the difference between two colors of adjacent vertices must not belong to $T$. In the latter, the vertices have to be partitioned into sets with specific pairwise distances, generalizing the concept of broadcast chromatic number~\cite{Goddard}. Other related approaches can be found in the \emph{weighted improper coloring}~\cite{Araujo} or the \emph{maximum contrast greyscale of graphs}~\cite{greyscale}.

We have already used spectrum graph coloring in different contexts such as Wireless
Surveillance Sensor Networks~\cite{de2015automated}, complex Wi-Fi networks~\cite{WCMCorthogonalchannels2018,hoz2017Wi-Fi,GDNcomplexnetwork2018}, or reconfiguration of a critical infrastructure network after a security incident~\cite{AAMAS-Resilience}. In the present paper we aim to face one of the biggest challenges in Wi-Fi channel assignment: developing a model which is complex enough to reflect all the conditions in the problem, but simple enough to allow the efficient computation of solutions. In particular, we show that our model can be applied to a real-world scenario with satisfactory results, both in terms of utility and of time consumption.

Section~\ref{sec:modelling} describes our network model and its simplification through edge contraction. Section~\ref{sec:experiments} motivates our approach, applies our model to the Wi-Fi network of the Polytechnic School of the University of Alcal\'a, and shows experiments to compare the detailed network model and the simplification using the \emph{de facto} standard technique for channel assignment and optimization via simulated annealing. Lastly, Section~\ref{sec:conclusions} outlines some concluding remarks, possible improvements and future lines of research.

\section{Modelling the Wi-Fi channel assignment problem}
\label{sec:modelling}

In this section, we use proximity graphs to model the wireless communication network, further using edge contraction to simplify the model in order to allow time-efficient optimization.

\subsection{Detailed model of a wireless communication network}
\begin{figure}[htb]
\centering
\includegraphics[width=\textwidth]{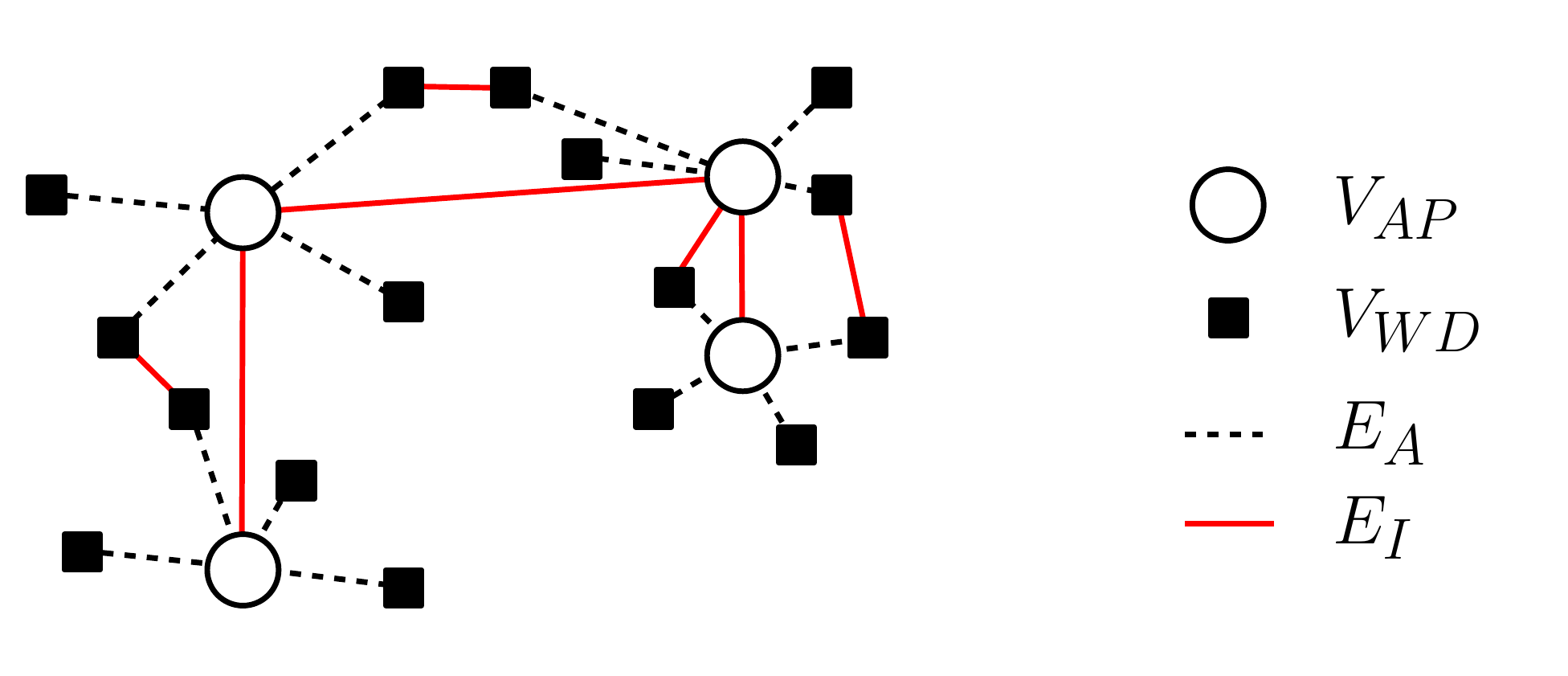}
\caption{Example of a network model.\label{fig:network}}
\end{figure}

We model the wireless communication network as a graph $G=(V,E)$ with two classes of vertices, $V=V_{AP}\cup V_{WD}$, and two classes of edges $E=E_{A}\cup E_{I}$. See Figure~\ref{fig:network}.

The vertices in $V_{AP}$ represent access points, devices acting as transmitters and receivers that connect wireless devices, represented by $V_{WD}$ vertices, to a network. The first class of edges, $E_{A}$, associates each wireless device to its closest access point, which is the one the device will use for communication. Hence, the edges in $E_{A}$ are those of the \emph{Bipartite Nearest Neighbor Graph} on $\{V_{AP}, V_{WD}\}$. The second class of edges, $E_{I}$, connects two vertices when their communications interfere. Since the communications among devices connected to the same access point are coordinated, wireless devices assigned to the same access point do not interfere to each other nor with that access point. Hence, the edges in~$E_{I}$ 
are a subgraph of the union of two \emph{Unit Disk Graphs} on $V_{AP}\cup V_{WD}$, for the interference distances appearing in~\cite{de2015automated}. The set $E_{I}$ is, more concretely, composed by edges that represent interferences between access points, between wireless devices not belonging to the same access points, or between an access point and a wireless device not belonging to it. Note that the vertices in $V_{WD}$ use the same channel as their associated access point, who handles the communications so that there is no interference between wireless devices associated to the same access point.

Thus, each channel assignment corresponds to a vertex coloring of~$V_{AP}$, with the 11 frequency channels $S=\{c_1,\ldots,c_{11}\}$ in the Wi-Fi spectrum corresponding to colors. In order to measure the utility of a coloring~$c$, we sum the utilities~$U_v$ of all the vertices~$v$ in the network. Such a \emph{detailed utility}~$U_v$ is defined normalizing the signal to noise ratio (\emph{SINR}), between the received signal~$P$ and the sum of received interferences~$I_{uv}$~\cite{Bazzi11}
\[
\mbox{\emph{SINR}}_v=\frac{P}{\sum_{u\in N(v)} I_{uv}},
\]
in the interval from~$0$ (corresponding to the value $\mbox{\emph{SINR}}_{min}$ below which the wireless devices cannot keep connected) to~$1$ (corresponding to the value $\mbox{\emph{SINR}}_{max}$ for which the throughput achieved by the wireless devices is the maximum that Wi-Fi technology permits). Thus, the detailed utility~$U_v$ is the normalized throughput at~$v$.

We use the values obtained in~\cite{Ng2012} for the interferences between Wi-Fi frequency channels, which constitute the matrix $W_{ij}=W(c_i,c_j)$ of interferences between colors in the Spectrum Coloring problem. However, this is only a part of the total interference received by a vertex. For two vertices $u,v$ using channels $c_i$ and $c_j$, the interference of the edge $uv$ has the form
\[
I_{uv} = W_{ij}+P_t+G_t+G_r-L-P_{loss}+\psi,
\]
where $P_t$ represents the transmission power (in dBm), $G_t$ (resp. $G_r$) stands for the transmission (resp. reception) antenna gain (in dB) and $L$ stands for losses due to walls, windows and other obstacles in the propagation (in dB). Further, the activity index $\psi$ accounts for higher bandwidth data flow occupying the wireless channel a higher fraction of the time. Finally, propagation losses (in dB) are reflected in $P_{loss}$, which we have computed according to~\cite{Green2002}
\[
P_{loss} = 7.6 + 40 \log_{10} d - 20 \log_{10} (h_t h_r),
\]
being $d$ the distance between interfering nodes, in meters, and being $h_t$ (resp. $h_r$) the transmission (resp. reception) antenna height, also in meters.
The interested reader can find more technical details in~\cite{hoz2017Wi-Fi}.

With the aforementioned description, we can formally formulate the problem as follows.
Given a geometric graph~$G$, together with a spectrum of~$k=11$ colors (channels) endowed with a $k\times k$ matrix~$W$ of interferences between them, the goal is to determine a $k$-coloring~$c$ for~$V_{AP}$ such that the sum of utility values for all the nodes in the graph (access points and wireless devices) is maximized, i.e., achieving the maximum
\[
\max\left\{\sum_{v\in V}U_v(G,W,c)\ \left|\right. \ c\mbox{ is a coloring of $V_{AP}$}\right\},
\]
where $U_v(G,W,c)$ denotes the detailed utility for the vertex~$v$ under the coloring~$c$, for the matrix~$W$ of interferences.

\subsection{Simplified models using edge contraction}
\label{subsec:contraction}

Clearly, the optimization problem just defined can be attacked by optimizers as, e.g., simulated annealing. However, the complexity of the model is expected to have great impact in the running times. This is why we aim to obtain a simplified model, which allows faster running times while obtaining good values of the detailed utility, i.e., good network throughputs.
In order to do so, we use edge contraction as follows:
\begin{enumerate}
    \item Inspired by the fact that wireless devices get the same color as the access point to which they are associated, we first contract all the edges in~$E_{A}$.
    \item Then, we assign a weight to the edges of the resulting graph $G\sslash E_{A}$, for which we have considered the following two possibilities:
    \begin{enumerate}
        \item A \textit{uniform contraction}, where parallel edges are merged and the resulting edge $uv$ gets a weight $w_{uv}=1$. While this loses most of the information from the original interference edges~$E_{I}$ and, hence, is expected to give worse results, it should also allow faster running times.
        \item A \textit{weighted contraction}, where parallel edges are merged and the resulting edge $uv$ gets a weight $w_{uv}$ equal to the number of merged edges. This reflects better the original interference edges~$E_{I}$, while still being much faster than evaluating all the interferences in~$E_{I}$.
    \end{enumerate}
\end{enumerate}

\begin{figure}[!htb]
    \centering
    \begin{subfigure}[b]{0.34\textwidth}
        \includegraphics[width=\textwidth]{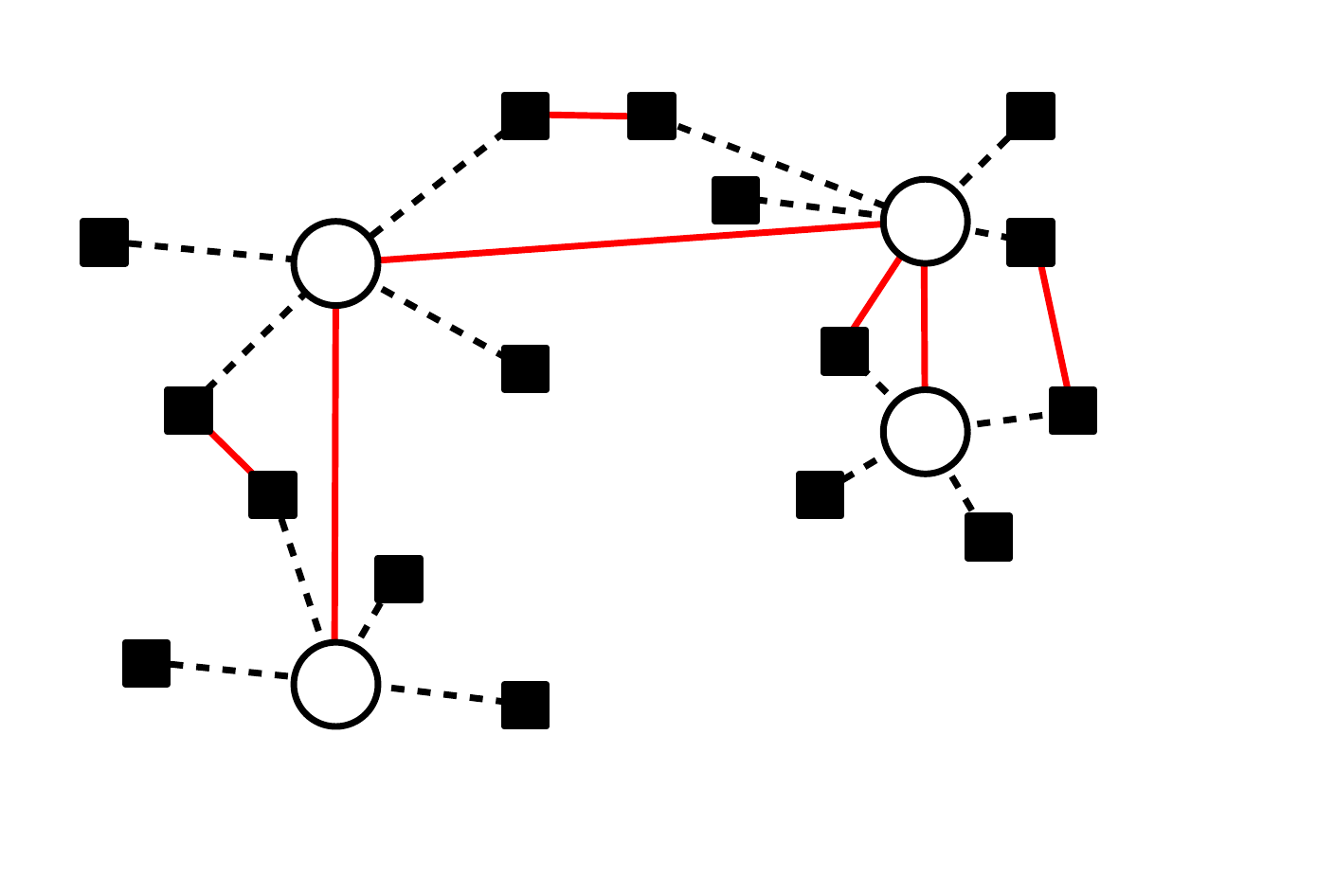}
        \caption{}
        \label{fig:jug}
    \end{subfigure}
    ~ 
    \begin{subfigure}[b]{0.3\textwidth}
        \includegraphics[width=\textwidth]{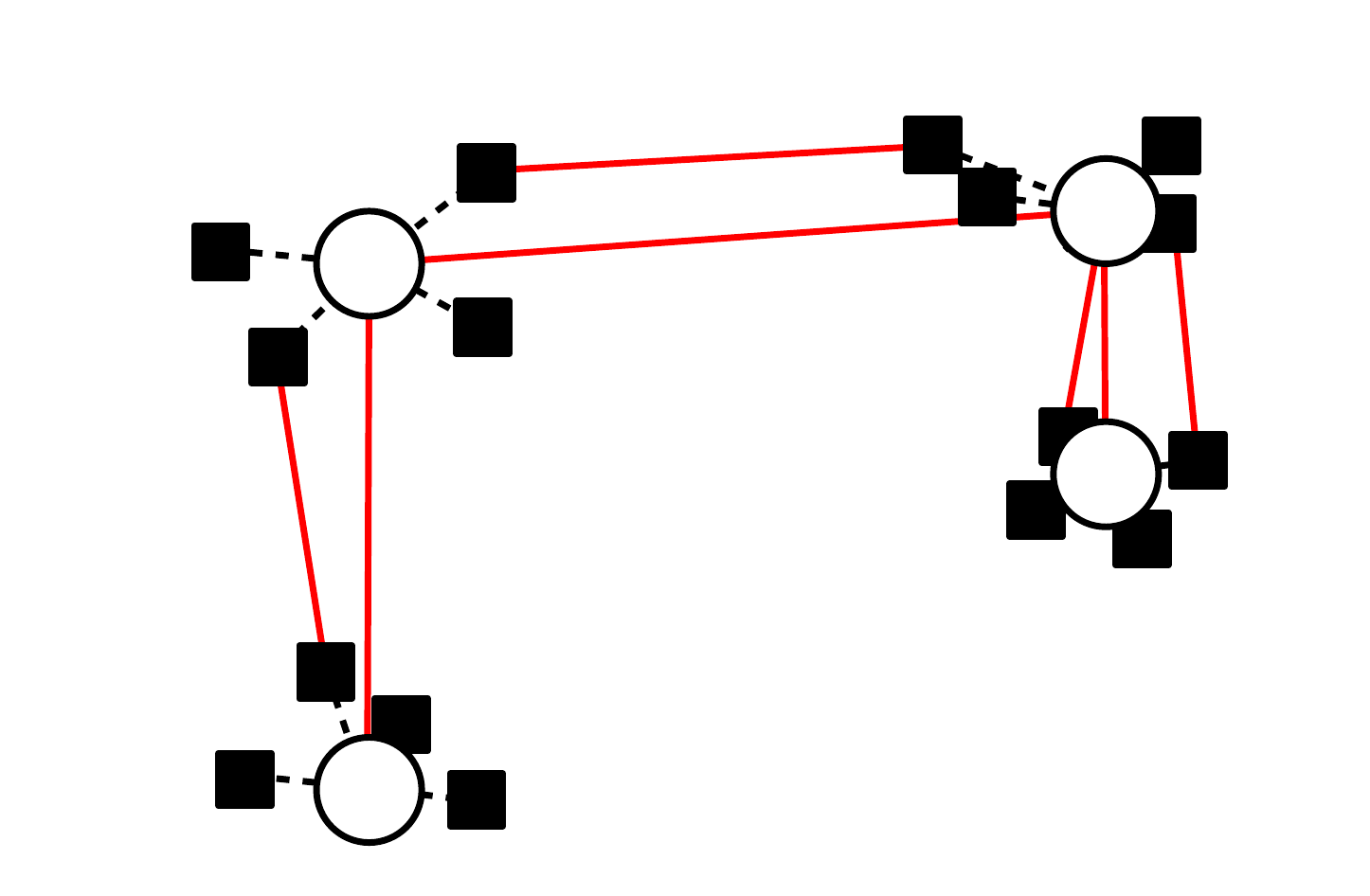}
        \caption{}
        \label{fig:contr}
    \end{subfigure}
    ~ 
    \begin{subfigure}[b]{0.3\textwidth}
        \includegraphics[width=\textwidth]{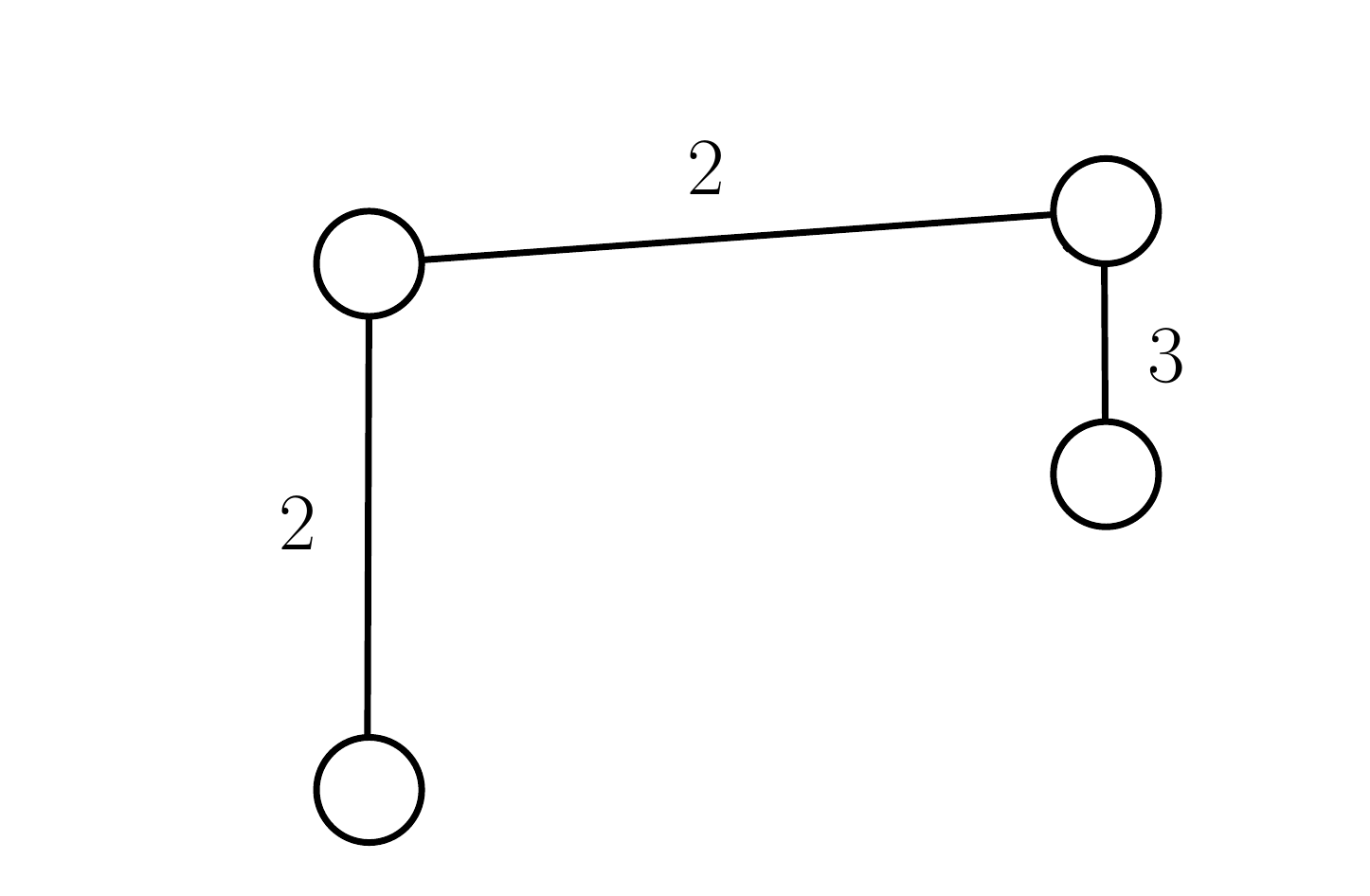}
         \caption{}
        \label{fig:weigh}
    \end{subfigure}
    \caption{Example of the simplification process. Figure~\ref{fig:jug} shows the network from Figure~\ref{fig:network}. Figure \ref{fig:contr} shows an intermediate step during the edge contraction of~$E_{A}$, for a better visualization, and finally Figure~\ref{fig:weigh} shows the resulting graph for the weighted contraction (for the uniform contraction $G\sslash E_{A}$, just remove the labels).}\label{fig:ejjuguete}
\end{figure}

Figure~\ref{fig:ejjuguete} shows an example of the contraction process, which can be checked for a real-world graph in Figure~\ref{fig:BIGgraph25}. Recall that circles symbolize access points and squares represent wireless devices, while dashed lines represent association to an access point and solid lines stand for interferences.

From a mathematical point of view, this simplification encodes in a single, weighted edge the number of interferences arising between two access points (APs) and the wireless devices (WDs) associated to them. This results in an abstract graph, which gets rid of the actual distances between interfering nodes and focuses on the connections between vertices in~$V_{AP}$, weighted according to how many interferences that connection is encoding.

Having simplified the graph, we need a \emph{simplified utility}~$U'_v$ of a vertex. Since the simplified graph has only weighted edges between vertices in~$V_{AP}$, it is natural to define the simplified utility of a vertex, for a coloring~$c$, as the weighted sum of the interferences induced by~$c$ at the edges incident to~$v$, i.e.,
\[
U'_v = \sum_{u\in N_{AP}(v)} w_{uv}W_{ij}
\]
where, as above, $W_{ij}$ is the interference between the colors $i=c(u)$ and~$j=c(v)$, and $w_{uv}$ is the weight of the edge~$uv$ in the simplified model.

\section{Experimental results}
\label{sec:experiments}

In order to test our approach, we have conducted two kinds of experiments. On one hand, we check to which extent good colorings for the simplified models are also good for the more detailed model. On the other hand, we compare several optimization techniques on the simplified and detailed models. We start by introducing the real-world setting we are dealing with.

\subsection{Real-world scenarios considered}

\begin{figure}[htb]
\centerline{
\includegraphics[width=0.55\columnwidth]{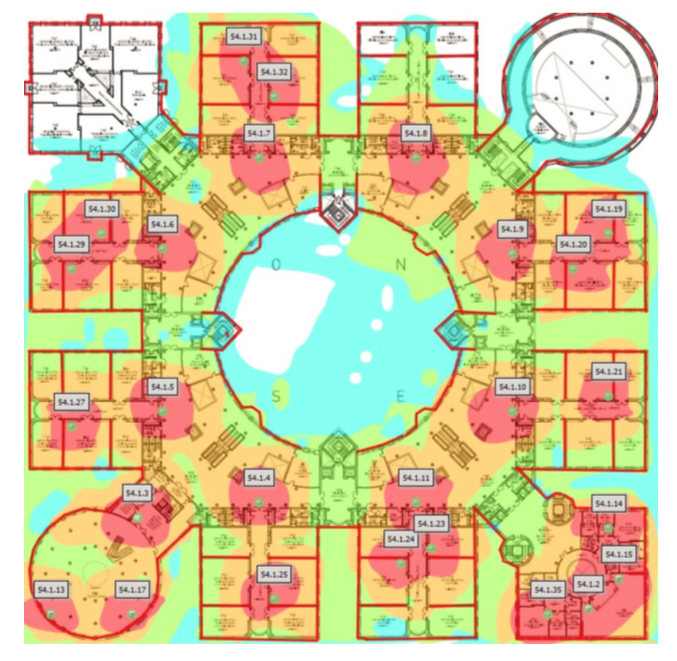}
}
\caption{Plan of the first floor of the Polytechnic School of the University of Alcal\'a.}
\label{fig:EPSplan}
\end{figure}

For a Wi-Fi real-world setting, we have used the actual positions of access points on the first floor of the Polytechnic School of the University of Alcal\'a, see Figure~\ref{fig:EPSplan}.

The building of the Polytechnic School has the shape of a square, with sides of approximately 130 meters long and a central courtyard. In the figure, the real positions of the $26$ access points are represented by green dots and the Wi-Fi signal coverage ranges from red (higher) to light blue (lower).

In order to position wireless devices on the map, we have considered $100$ students randomly located in this first floor (resting, studying...) and also users attending lectures in classrooms. In order to do so, we have tested several scenarios varying randomly the ratio~$\rho$ of classrooms being used, to reflect the different occupation rates during an academic day. In particular, we have considered occupation ratios $\rho \in \{0.25, 0.5, 0.75, 1.0\}$. With a total of $48$ classrooms in this floor of the building, this means that we have considered scenarios with, respectively, $12$, $24$, $36$ and~$48$ classrooms being used at the same time.

Further, at each classroom we have considered a number of~$25$ students, which is the standard in this floor since the classrooms are laboratories and, hence, the groups have a limited number of students. We have deployed these $25$ students in the classroom at random, using a normal distribution around the center of the room and a standard deviation, normalized to the size of the scenario, of~$0.05$. Finally, as the specific random classrooms under use could affect the results, we have tested three sets of experiments for each value of $\rho$, giving a total number of~$12$ scenarios under study, with ten executions of the experiment for each of them.

\begin{figure}[htb]
    \centering
    \begin{subfigure}[b]{0.45\textwidth}
        \includegraphics[width=\textwidth]{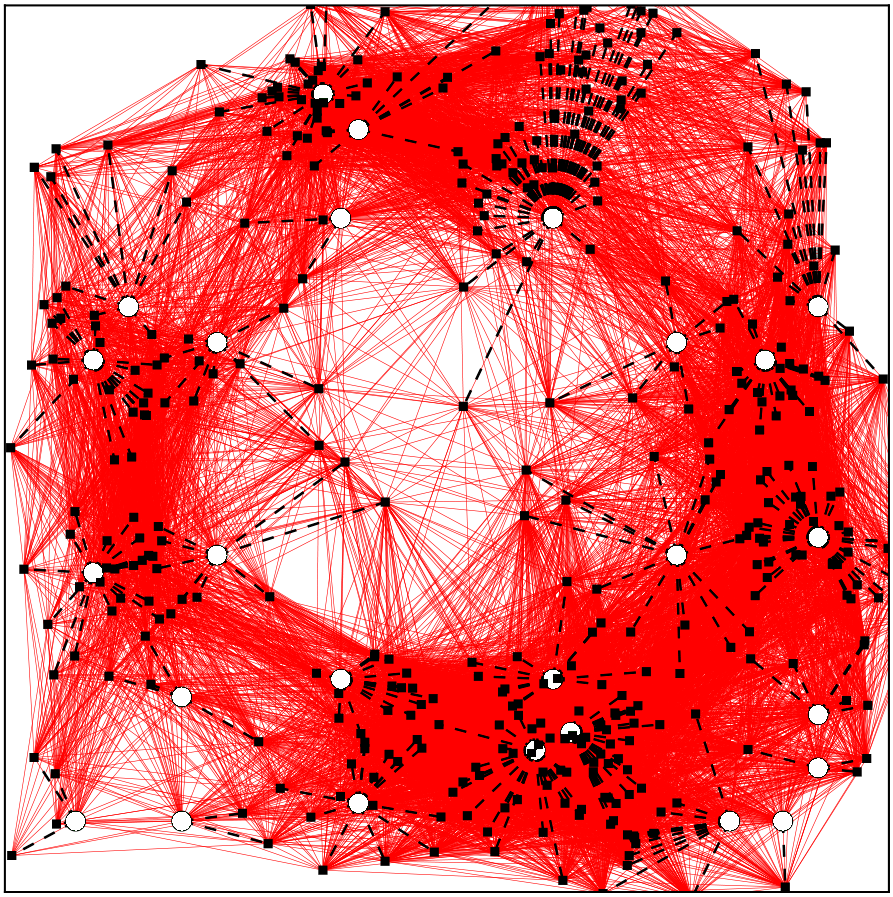}
        \caption{}
        \label{fig:bnngudg}
    \end{subfigure}
    ~ 
    \begin{subfigure}[b]{0.45\textwidth}
        \includegraphics[width=\textwidth]{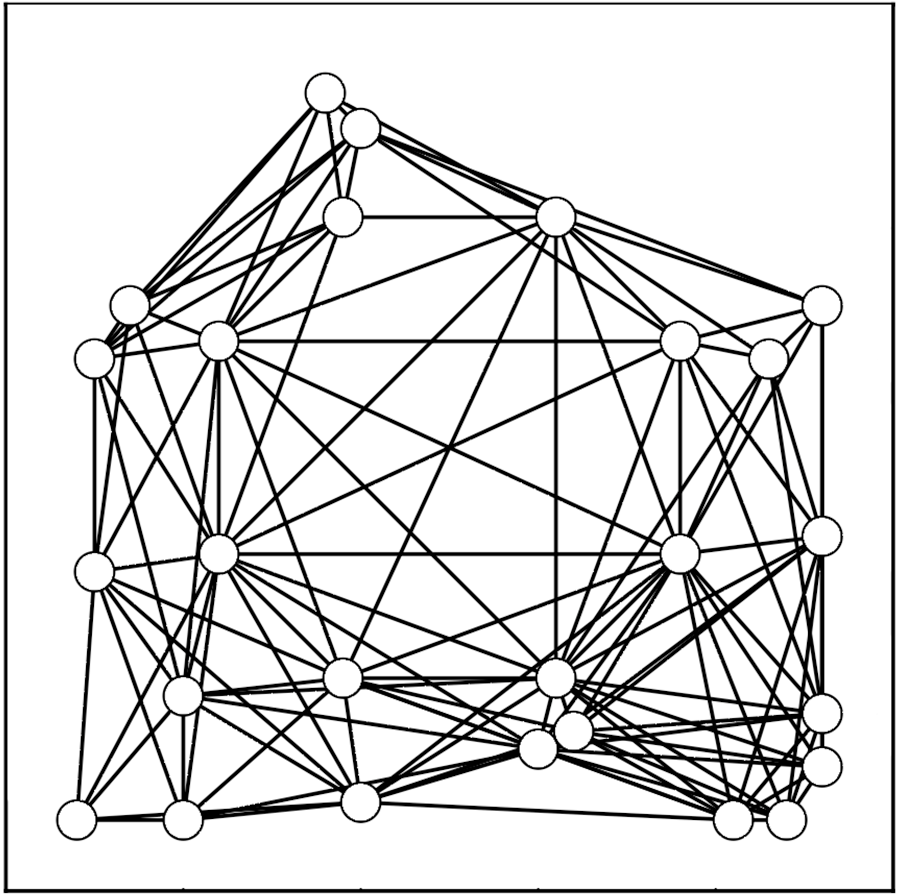}
         \caption{}
        \label{fig:simpl25}
    \end{subfigure}
    \caption{Example of the graphs for one of the simplest real-world scenarios.
    (a) Whole graph $G=(V_{AP}\cup V_{WD},E_A\cup E_I)$, following the legend in Figure~\ref{fig:ejjuguete}.
    (b) Uniform contraction~$G\sslash E_{A}$.}\label{fig:BIGgraph25}
\end{figure}

Figure~\ref{fig:BIGgraph25} allows the reader to apprehend the high complexity of the network, even for the simplest scenarios. The whole graph $G= (V_{AP}\cup V_{WD},E_A\cup E_I)$ is shown in Figure~\ref{fig:BIGgraph25}a for one of the three scenarios with only a~$25\%$ of the classrooms being used, while Figure~\ref{fig:BIGgraph25}b shows the simplification obtained after performing the uniform contraction~$G\sslash E_{A}$.

\subsection{Correlation between the detailed and simplified utilities}
\label{subsec:GoodColorings}

In order to check to which extent good solutions of the simplified model are also good for the detailed model, we have considered two types of graphs. First, of course, the realistic graphs presented in the previous subsection.
Second, we have considered randomly generated scenarios with different sizes of~$|AP|$ and~$|WD|$. In particular, we have created scenarios with 15, 50 and 100 access points considering, for each of these values, three different settings with 1, 5, and 10 wireless devices per access point. Thus, with the
notation $(|AP|,|WD|)$, we have considered the following combinations: (15, 15), (15, 75), (15, 150), (50, 50), (50, 250), (50, 500), (100, 100), (100, 500), and (100, 1000). For each of these 9 combinations, we have generated 30 different random graphs, resulting in a total of 270 graphs. These cover a reasonable range of scenarios, not only in terms of the
problem size, from a few tens to more than one thousand of devices, but also in terms of variability, due to randomness.

For each of the graphs considered, we have computed 1000 different random colorings, and for each of this colorings:
\begin{itemize}
   \item We have computed, for the considered graph~$G$ and coloring~$c$, the average~$\overline{U_v}$ of the detailed utilities at vertices~$v$.
   \item For the graph~$G$ we have obtained the uniform and weighted contractions, $G_u$ and~$G_w$ as in Subsection~\ref{subsec:contraction}. For both simplifications, we have computed the average~$\overline{U'_v}$ of the simplified utilities at vertices~$v$ under the coloring~$c$ considered.
\end{itemize}

\begin{figure}[htbp]
    \centering
    \begin{subfigure}[b]{0.49\textwidth}{\adjustbox{trim={.18\width} {.03\height} {0.2\width} {.01\height},clip}  {\includegraphics[width=1.6\textwidth]{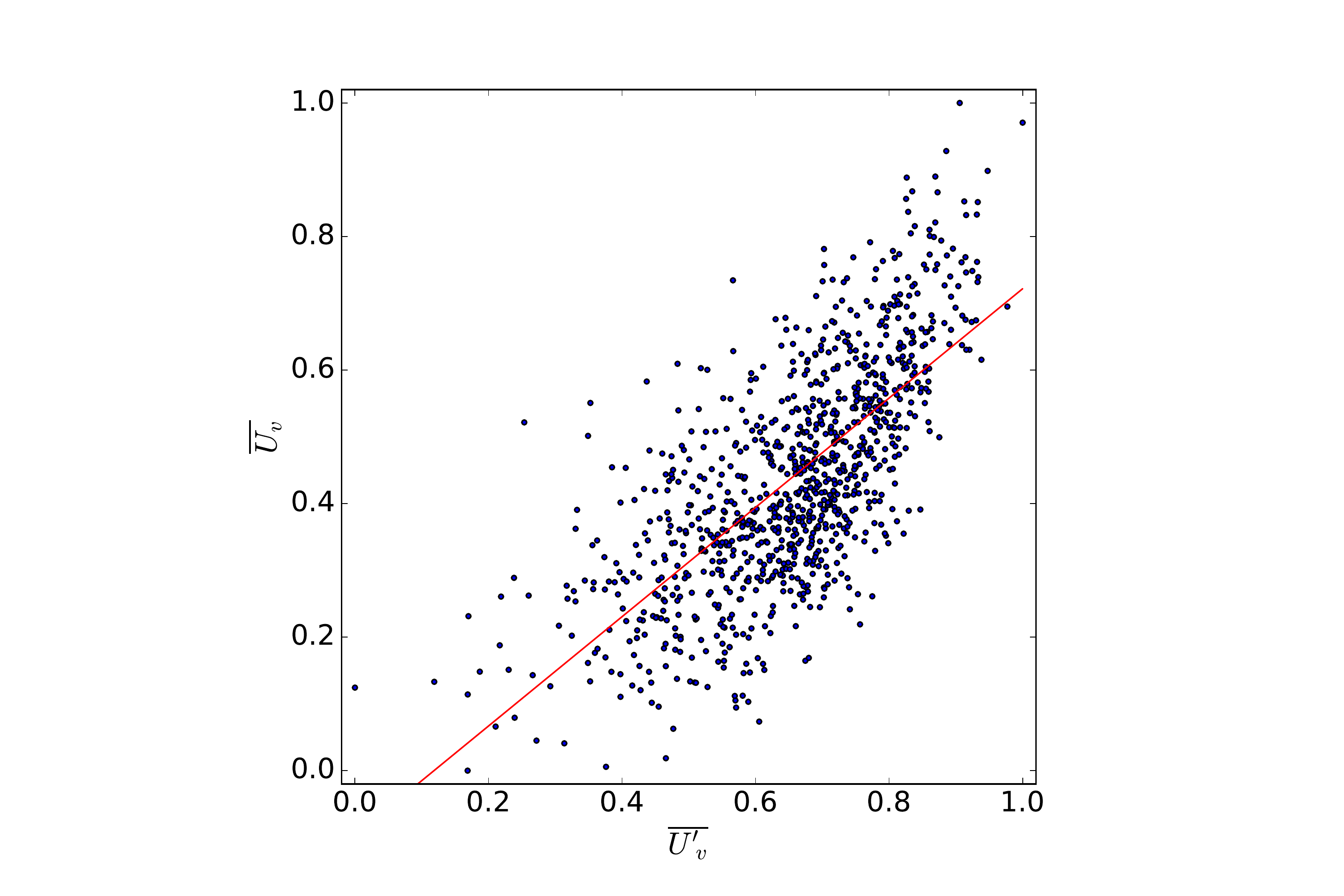}}}
        \caption{}
        \label{fig:EPS-025-3_1ut_uc}
    \end{subfigure}
    \begin{subfigure}[b]{0.49\textwidth}{\adjustbox{trim={.18\width} {.03\height} {0.2\width} {.01\height},clip}  {\includegraphics[width=1.6\textwidth]{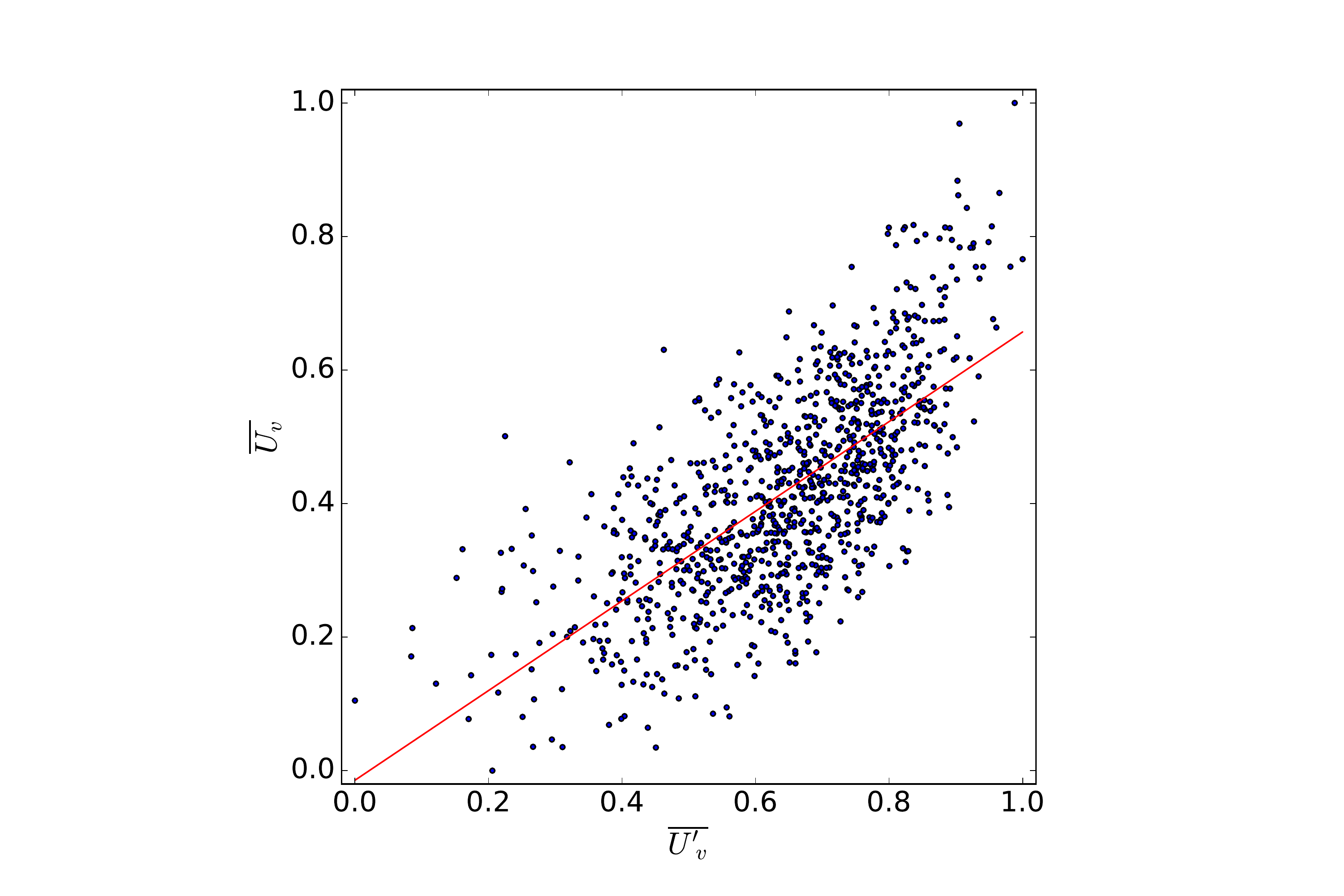}}}
        \caption{}
        \label{fig:EPS-075-3_1ut_uc}
    \end{subfigure}
    \\
    \begin{subfigure}[b]{0.49\textwidth}{\adjustbox{trim={.18\width} {.03\height} {0.2\width} {.01\height},clip}  {\includegraphics[width=1.6\textwidth]{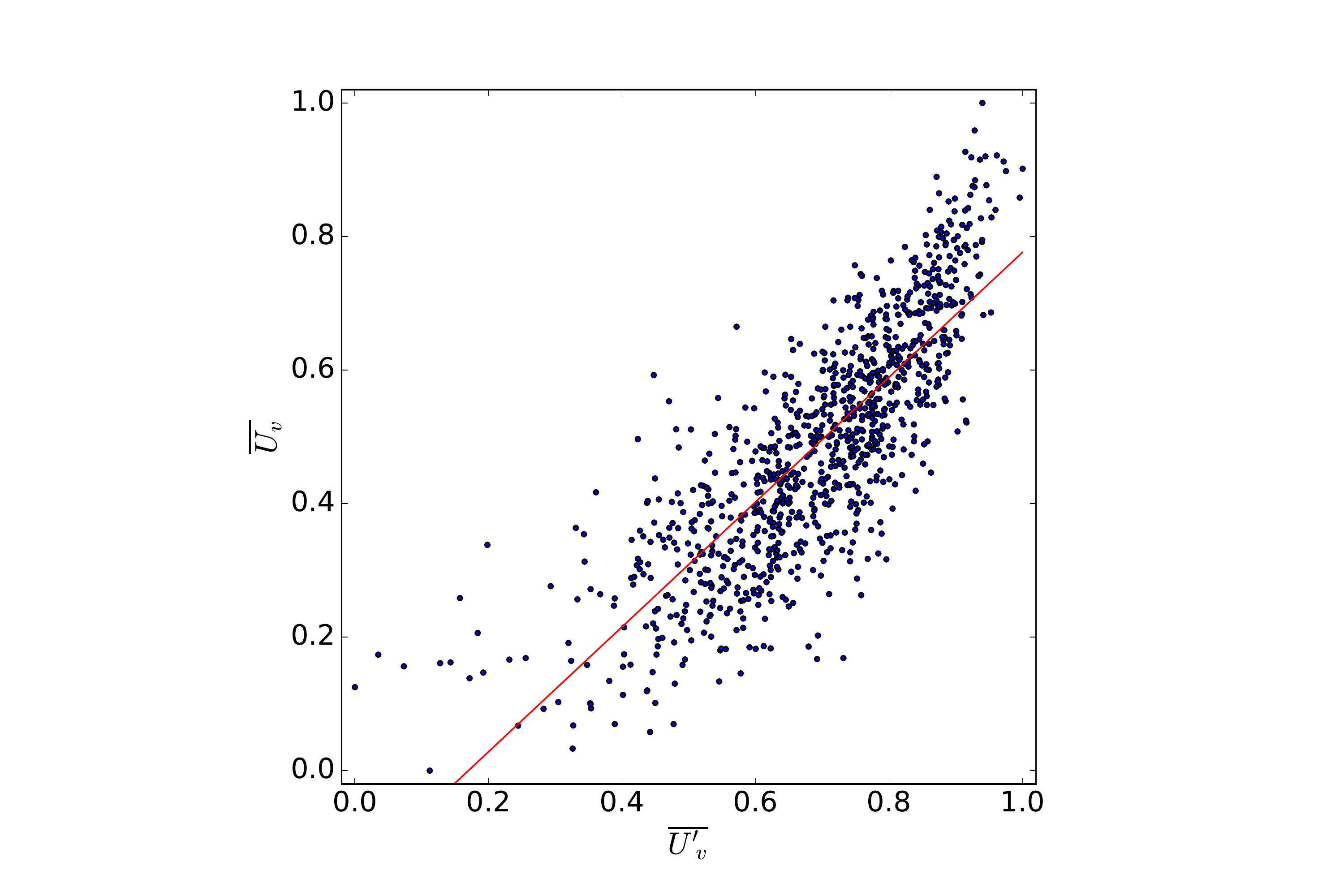}}}
        \caption{}
        \label{fig:random-random-4-10-10ut_uc}
    \end{subfigure}
    \begin{subfigure}[b]{0.49\textwidth}{\adjustbox{trim={.18\width} {.03\height} {0.2\width} {.01\height},clip}  {\includegraphics[width=1.6\textwidth]{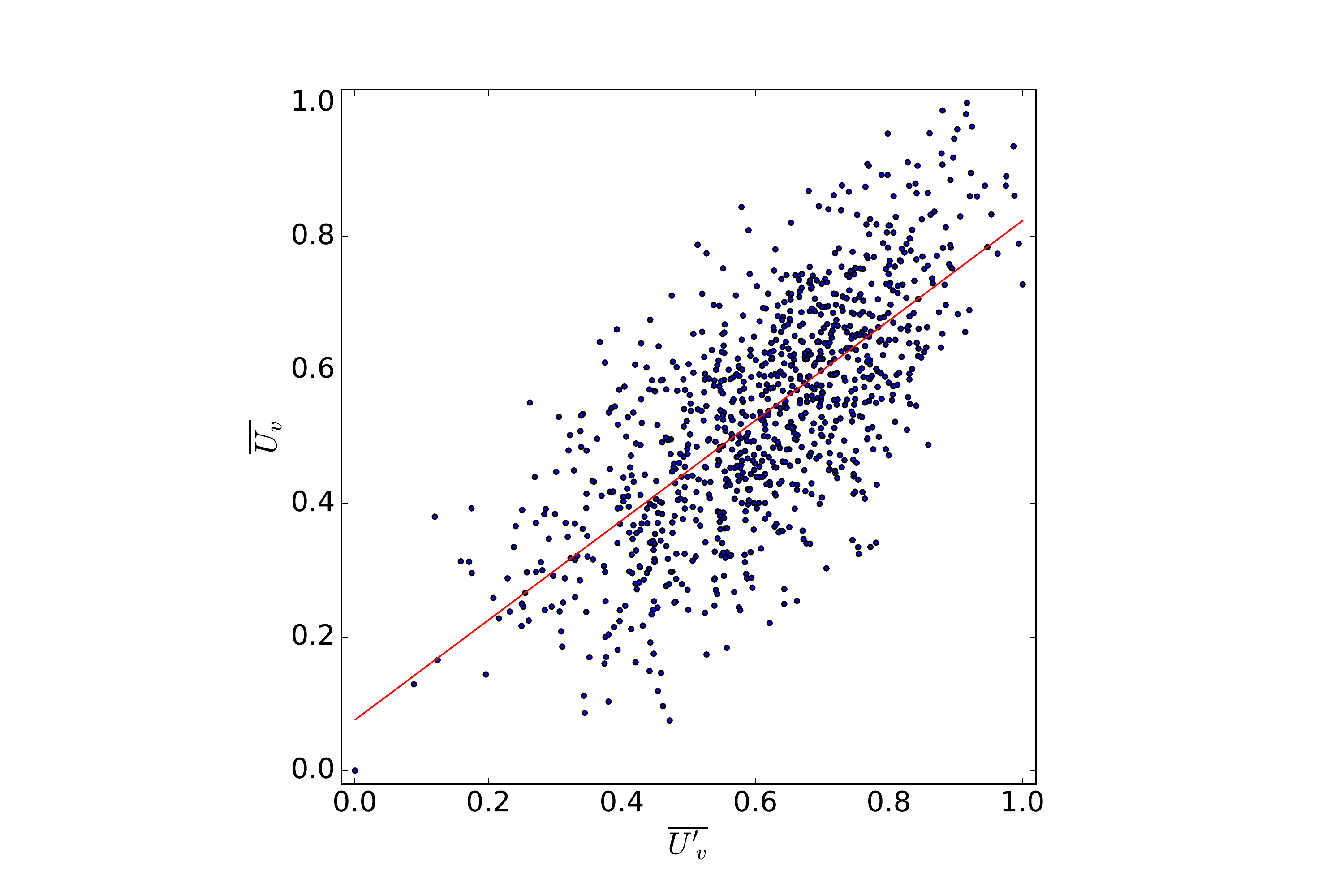}}}
        \caption{}
        \label{fig:random-random-10-1-19ut_uc}
    \end{subfigure}
\caption{Comparison of average utilities~$\overline{U'_v}$ and $\overline{U_v}$ for two realistic (top) and two random (bottom) graphs.}
\label{fig:ComparisonGoodColorings}
\end{figure}

Figure~\ref{fig:ComparisonGoodColorings} shows the comparison of these average utilities~$\overline{U_v}$ and $\overline{U'_v}$ (for the weighted contraction $G_w$), after normalizing from~0 (average utility in the worst case) to 1 (that in the best case), for two graphs corresponding to the realistic setting, subfigures (a)-(b), and for two graphs corresponding to random scenarios, subfigures (c)-(d). Each of the 1000 random colorings is depicted as a point, whose $x$-coordinate, respectively $y$-coordinate, is the average simplified utility~$\overline{U'_v}$, respectively average detailed utility~$\overline{U_v}$. The regression line is included.

\begin{figure}[htb]
    \centering
    \includegraphics[width=\textwidth]{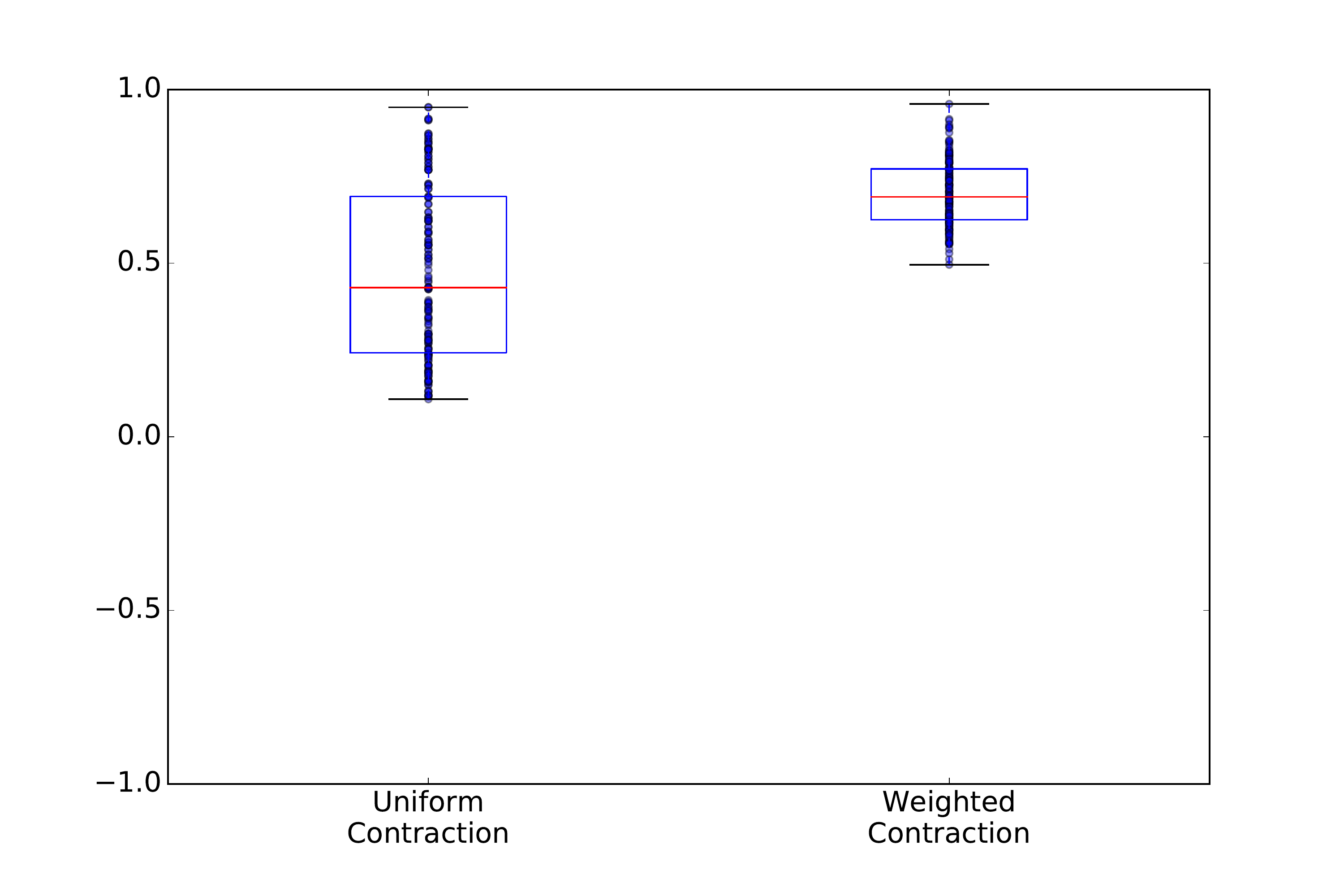}
    \caption{Box plots for the Pearson correlation coefficients for the experiments in Subection~\ref{subsec:GoodColorings}.}
\label{fig:BoxPlots}
\end{figure}

The tendency shown in these four graphs is that good colorings for the simplified model, i.e., good values of~$\overline{U'_v}$, turn out to be also good colorings for the detailed model, i.e., good values of~$\overline{U_v}$. For the sake of checking that this tendency is common to all the~282 graphs considered, Figure~\ref{fig:BoxPlots} shows the box plots for the Pearson correlation coefficients for all the experiments in this section, using either the uniform contractions~$G_u$ (left) or the weighted contractions~$G_w$ (right).

Results show that both contractions exhibit a positive correlation with respect to the detailed utility. Comparing both contractions, we conclude that, as expected, the correlation of the weighted contraction with the detailed utility is much higher than the one with the uniform contraction.

\subsection{Optimization techniques considered}

We have chosen as benchmark the \emph{de facto} standard for Wi-Fi channel assignment, called Least Congested Channel Search (LCCS)~\cite{Achanta}. In this technique each access point scans the channel occupation and, asynchronously, chooses the channel which minimizes the interferences from other active access points and their associated wireless devices. We have implemented a coordinated LCCS, which is the choice usually implemented in corporate environments, where a centralized controller evaluates the changes proposed before their implementation, in order to prevent utility oscillations.

Against this standard, we propose the combination of our network models together with a simulated annealing (SA) optimizer~\cite{Kirkpatrick83}. Our choice of this metaheuristic relies in the fact that it is one of the best known and most widely used optimizers, having a large number of implementations and being reasonably easy to use.
We have compared the performance of SA for three different types of inputs. First, for the whole graph $G=(V_{AP}\cup V_{WD},E_A\cup E_I)$, hence having all the information available, for which we will use the notation~$SA_g$. Second, for the uniform contraction~$G\sslash E_{A}$, with the notation~$SA_u$. Finally, for the weighted contraction which adds weights to the edges of~$G\sslash E_{A}$ as explained in Subsection~\ref{subsec:contraction}, which we denote by~$SA_w$.

Note that, while $SA_g$ aims to find a coloring which optimizes the detailed utility~$\sum_{u\in V} U_v$, both $SA_u$ and~$SA_w$ aim to find a coloring which optimizes the simplified utility~$\sum_{u\in V_{AP}} U'_v$. In these two latter cases, the coloring obtained is afterwards applied to the whole graph~$G$, and the induced detailed utility is computed. Thus, in the following subsection we can compare the results for the detailed utility, or normalized throughput.

\subsection{Results obtained}

\begin{figure}[htb]
    \centering
        \includegraphics[width=\textwidth]{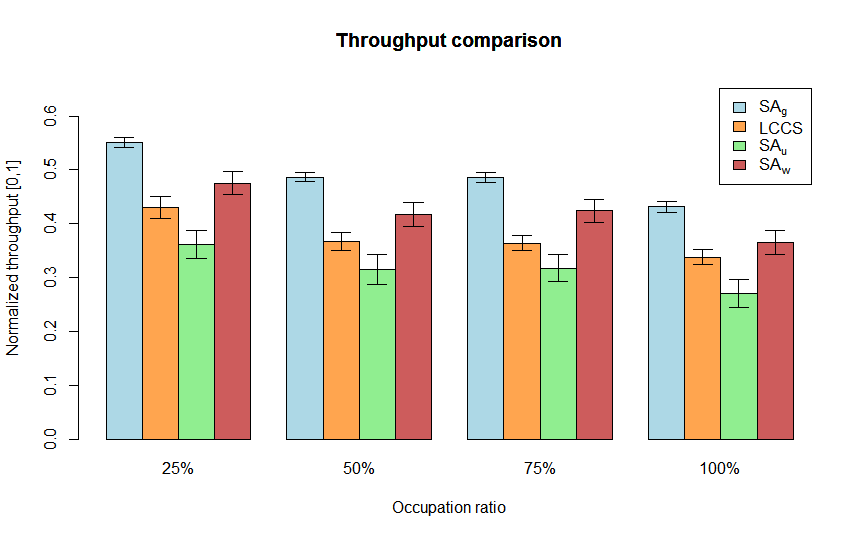}
        \caption{Bar graphics for the utility (throughput) obtained by the four alternatives considered.}
        \label{fig:utility_results}
\end{figure}

\begin{figure}[htb]
        \includegraphics[width=\textwidth]{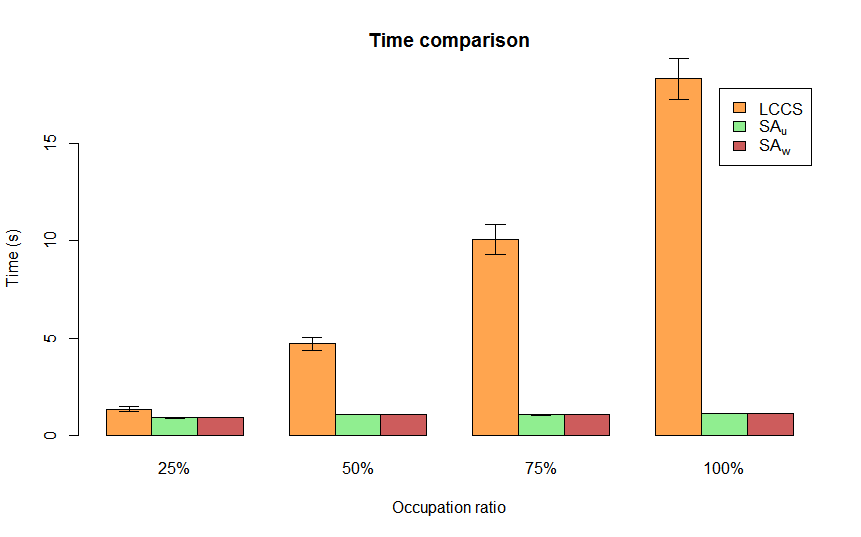}
        \caption{Bar graphics for the time consumed by three of the four alternatives considered. (Times for $SA_g$ are several orders of magnitude higher.) }
        \label{fig:time_results}
\end{figure}

We have conducted these experiments on an Intel\textregistered\  Core\textsuperscript{\texttrademark} i7-2600 with
8 CPUs@3.40GHz and 8GB RAM, running Ubuntu 14.04.4 LTS. Table~\ref{tab:results_eps} shows the average results obtained for the four alternatives considered, $SA_g$, LCCS, $SA_u$, and $SA_w$, concerning the average detailed utility (normalized throughput) per vertex and the computation time. Note that this average is the result of dividing the sum of utilities $\sum_{v \in V}U_v$ by the number of nodes, which is constant for each graph.
These data are graphically represented in Figures~\ref{fig:utility_results} and~\ref{fig:time_results}, including the $95\%$ confidence intervals.

\begin{table}[tb]
\centering
\caption{Evaluation of the four alternatives considered, in terms of average detailed utility per vertex, in~$[0,1]$, and computation times, in seconds. The parameter $\rho$ denotes the ratio of classrooms being used.}
\label{tab:results_eps}
\renewcommand{\arraystretch}{0.95}
\setlength{\tabcolsep}{.13cm}
\begin{tabular}{| c | c c | c c | c c | c c |}
\hline
 \rule{0pt}{10pt}
\multirow{2}{*}{$\rho$} & \multicolumn{2}{|c}{$SA_g$} & \multicolumn{2}{|c}{LCCS} & \multicolumn{2}{|c}{$SA_u$} & \multicolumn{2}{|c|}{$SA_w$}\\
\cline{2-9}
 \rule{0pt}{11pt}
 & $\overline{U_v}$ & Time & $\overline{U_v}$ & Time & $\overline{U_v}$ & Time & $\overline{U_v}$ & Time \\
\hline
\multirow{3}{*}{0.25} & 0.558 & 691.75 & 0.432 & 1.33 & 0.383 & 1.00 & 0.472 & 1.02\\
& 0.532 & 668.63 & 0.405 & 1.38 & 0.319 & 0.80 & 0.458 & 0.81\\
& 0.564 & 738.69 & 0.454 & 1.37 & 0.382 & 0.91 & 0.497 & 0.91\\
\hline
\multirow{3}{*}{0.5} & 0.468 & 2182.68 & 0.339 & 5.16 & 0.306 & 1.02 & 0.405 & 1.03\\
& 0.493 & 2055.25 & 0.378 & 3.85 & 0.326 & 0.96 & 0.405 & 0.96\\
& 0.499 & 2021.83 & 0.384 & 5.13 & 0.314 & 1.22 & 0.444 & 1.24\\
\hline
\multirow{3}{*}{0.75} & 0.466 & 3927.16 & 0.360 & 10.12 & 0.304 & 1.03 & 0.405 & 1.05\\
& 0.519 & 4090.70 & 0.368 & 10.56 & 0.330 & 1.05 & 0.448 & 1.08\\
& 0.474 & 3902.17 & 0.365 & 9.58 & 0.318 & 1.10 & 0.419 & 1.12\\
\hline
\multirow{3}{*}{1} & 0.427 & 6683.11 & 0.338 & 18.54 & 0.277 & 1.17 & 0.356 & 1.19\\
& 0.444 & 6502.87 & 0.345 & 16.26 & 0.286 & 1.07 & 0.381 & 1.08\\
& 0.425 & 6280.99 & 0.331 & 20.21 & 0.249 & 1.16 & 0.358 & 1.18\\
\hline
\end{tabular}
\end{table}

    ~ 

The results show that the best performance in terms of the average utility for each node is clearly obtained by the simulated annealing optimizer being applied to the whole graph~$G$, since all the information of the network is available in this case ($SA_g$). However, the computation time is very high for this alternative, actually several orders of magnitude higher than for the others. Note that for all the techniques under study we have used the same number of iterations, concretely~$3000$.

On the contrary, applying $SA$ to the contractions we propose in this work, $SA_u$ for the uniform contraction and $SA_w$ for the weighted contraction, allows much lower computation times. The throughput obtained by these two alternatives is, of course, worse than the one for the whole graph. Nevertheless, the throughputs obtained for the weighted contraction $SA_w$ are not only better than the ones for the uniform contraction $SA_u$ (which is natural because of reflecting more information), but also they clearly beat the \emph{de facto} standard LCCS.

Actually, our proposal $SA_w$ outperforms LCCS in all cases, not only in terms of throughput, but also in terms of computation time, with the difference being larger for the more complex scenarios in which more classrooms are in use. In conclusion, the techniques we propose lead to better and faster results than the real-world standard, using a simplified model that avoids working with the whole graph and which would allow to optimize Wi-Fi channel assignments much closer to real time.

\section{Concluding remarks}
\label{sec:conclusions}
We have proposed a graph model $G=(V_{AP}\cup V_{WD},E_A,E_I)$ for Wi-Fi networks with two types of vertices, corresponding to access points and wireless devices, and two types of edges, corresponding to associations to access points and interferences. These edges come from the use of proximity graphs, more concretely a Bipartite Nearest Neighbor Graph and two Unit Disk Graphs. We have further modeled interferences by considering a Spectrum Graph Coloring problem, enriched to better reflect the real conditions.

Inspired by the associations in $E_A$, of each wireless device to an access point, we have proposed a simplification of the graph via edge contraction, $G\sslash E_{A}$. In addition to this uniform contraction, we have considered the weighted contraction which assigns to each edge the number of edges contracted on it.

We have performed experiments checking to which extent a good coloring of the simplified graph is also good for the original graph. Finally, we have performed experiments to compare the \emph{de facto} standard for Wi-Fi channel assignment, LCCS, to a simulated annealing optimizer applied to the whole graph, $SA_g$, to the uniform contraction, $SA_u$, and to the weighted contraction, $SA_w$. As expected because of having all the information available, the best throughput is obtained by $SA_g$. However its computation time is several orders of magnitude higher than that of the other three alternatives. On the contrary, $SA_w$ beats not only the simplest alternative~$SA_u$ but also the real-world standard LCCS, giving better and faster results.

This work opens a number of new opportunities for further research. First, we would like to study whether other simplifications of the network model might provide an even better balance between the quality of the throughput and the computation time. An alternative approach might be to keep the full network model and apply the simplifications devised in this paper (or new ones) to the very process of neighbor generation in the simulated annealing algorithm, using a similar methodology to the one in~\cite{duque1997constructing}. Direct application of the methodology in that paper is not feasible, since our problem does not have hard compatibility constraints. However, a similar approach could be devised deriving ``softer'' constraints which could help us not to reduce the search space (we do not have \textit{a priori} knowledge of what an optimal solution looks like), but to guide the search through the whole space in order to maximize the probability of finding optimal solutions.

\section{Acknowledgements}
The authors have been partially supported by H2020-MSCA-RISE project 73499 - CONNECT, by MINECO projects MTM2014-54207, and TIN2016-80622-P (AEI/FEDER, UE), and by the Spanish Ministry of Science project MTM2017-83750-P (AEI/FEDER, UE). We would also like to acknowledge the valuable suggestions from anonymous referees which helped to improve the paper.

\end{document}